\documentclass[12pt,a4paper]{article} 
\usepackage[utf8]{inputenc}
\usepackage{braket}
\usepackage[english]{babel}
\usepackage{a4wide}
\usepackage{graphicx}
\usepackage{amssymb}
\usepackage{amsmath}
\usepackage{textcomp}
\usepackage{caption}
\usepackage{float}
\usepackage{wrapfig}
\usepackage{subcaption}
\usepackage[dvipsnames]{xcolor}
\usepackage{mathrsfs}
\usepackage{tikz}
\usepackage{authblk}
\usepackage[authoryear]{natbib}
\usepackage[hidelinks]{hyperref}

\setcitestyle{authoryear, open={(},close={)}}

\def\dd{\mbox{d}}
\newcommand{\sm}[1]{\mbox{\scriptsize #1}}
\newcommand{\tn}[1]{\mbox{\tiny #1}}

\title{\bf Supersymmetry in the Seiberg-Witten Theory: A Window into Quantum Field Theory}
\author[1]{Sanne Vergouwen}
\author[2]{Sebastian De Haro}
\affil[1]{\small Freudenthal Institute, Utrecht University}
\affil[2]{\small Institute for Logic, Language and Computation and Institute of Physics, University of Amsterdam}

\date{\today}

\begin{document}

\maketitle


\begin{center}
\textbf{\large \bf Abstract}
\end{center}

We take supersymmetry in the Seiberg-Witten theory as a case study of the uses of (super)symmetry arguments in studying the ontology of four-dimensional interacting quantum field theories. Together with a double expansion, supersymmetry is a via media that helps to bridge the gap between the ontologies of an exact quantum field theory and its semi-classical limit.
We discuss a class of states that exist at any value of the coupling, and whose properties such as mass, electric and magnetic charges, and spin quantum numbers can be precisely characterised at low energies. The low-energy theory is best presented as a one-dimensional complex manifold, equipped with metric and other structures: namely, the space of low-energy vacua, covered by three open regions that are interpreted as macroscopic phases. 
We discuss two cases of emergence: the emergence of the low-energy regime and the emergence between models at low energies, thereby highlighting the significance of the topology of the space of vacua for such cases of emergence.

\vskip 1truecm
\noindent
Forthcoming in {\it Synthese}'s Topical Collection `Establishing the Philosophy of Supersymmetry'.

\thispagestyle{empty}
\addtocounter{page}{-1}

\newpage

\tableofcontents
\section{Introduction}\label{intro}

A major challenge in the study of quantum field theories is bridging the gap between the ontologies of an exact theory and the ontology of its semi-classical limit, where the latter is an approximation of the physics of that theory. In realistic interacting quantum field theories, it is in general difficult to calculate physical quantities by using exact methods. Thus a common approach is to study the {\it perturbative regime} of such quantum field theories, where an exact solution is approximated by perturbation theory. However, perturbative theories, being approximative,  lack the mathematical rigour that many philosophers hope for when they interpret quantum field theories. 

In this paper, we argue that there is a {\it via media} for studying the ontology of interacting quantum field theories. This via media is a middle way between the exact theory and the semi-classical limit. Thus, although it is not an exact treatment of the high-energy regime, it does, by its giving a fully non-perturbative treatment of the strong coupling effects, go well beyond the semi-classical limit. It thereby establishes a link between the properties of the theory's states and quantities at weak, and at strong, coupling. Furthermore, our application of this via media focusses on interacting four-dimensional supersymmetric quantum field theories that show behaviours, such as the Higgs mechanism and quark confinement, that are very similar to those of the standard model of particle physics. Thus we will argue that, given the difficulty of bridging the gap between the ontologies of an exact theory and that of its semi-classical limit, philosophers would benefit from taking notice of this approach.

We take the Seiberg-Witten theory, a four-dimensional supersymmetric quantum field theory, as a case study. This is a low-energy theory that is not meant to provide an actual description of real-world phenomena. Rather, it is a toy theory that helps to improve our understanding of phenomena, such as confinement, that we find in more realistic quantum field theories. It was named after Nathan Seiberg and Edward Witten, who in 1994 succeeded in writing down an exact analytic solution of it at low energies. A key feature of the Seiberg-Witten theory is that the state-space of the low-energy regime is a manifold that is a topologically non-trivial Riemann surface.\footnote{A Riemann surface is a one-dimensional connected complex manifold. Here, it will come endowed with a positive-definite metric and a symplectic structure.} The complete theory is given by three low-energy models defined on this manifold that are related to each other by dualities. Each model is characterised by the expectation value of a particular field (e.g.~the Higgs field). The expansion of the effective action in terms of this field has a limited radius of convergence. Where the series does not converge, it can be resummed so that a new effective action, based on the expectation value of a dual field, can be found.\footnote{We will explain the sense in which we use the word `dual' in Section \ref{SWth}.}

The via media that we advocate has two main ingredients, both of which are present in the Seiberg-Witten theory: the first is supersymmetry; the second is a double expansion. Supersymmetry is a spacetime symmetry that extends the Poincar\'e spacetime symmetries. Like other symmetries in quantum field theory, supersymmetry sets restrictions on the form of the action. However, supersymmetry is particularly successful in constraining the dynamics of a theory to the extent that it is possible to understand it analytically. This analyticity makes it a valuable symmetry in the study of non-perturbative regimes of quantum field theories, where the semi-classical approximation cannot be used reliably.\footnote{One could of course adopt other simplifying methods or approaches to study realistic phenomena in quantum field theory, such as considering lower spacetime dimensions: and such studies have indeed been done. However, the drawback of lower-dimensional models is that we can see no way of experimentally verifying them, while supersymmetry is, in principle, verifiable. Furthermore, lower-dimensional models are qualitatively very different (much simpler) than four-dimensional ones, while the kinds of effects that we find in supersymmetric theories resemble qualitatively, if not quantitatively, the physics of the real world.} For example, there exist supersymmetric accounts that have demonstrated confining behaviour analytically, and that explain it as the result of monopole condensation.

The second ingredient, i.e.~the double expansion, is not necessarily limited to supersymmetric quantum field theories, and it can be used more generally. The idea is that for theories with scalar fields and non-trivial potentials, such as the Higgs field, the expectation value of the scalar field (usually forming a dimensionless combination with a constant parameter, such as a cut-off for the momentum) can be used as an expansion parameter, analogous to a coupling constant. There are then {\it two} expansions, each correcting the Lagrangian in its own sort of way. These two types of corrections are: 

(i) corrections in the expectation value of the scalar field that renormalize the coupling functions in the Lagrangian, and 

(ii) corrections by high-energy modes that contribute higher-dimensional operators to the Lagrangian.

This combination of supersymmetry and double expansion as via media uses the fact that, although supersymmetry cannot guarantee exactness with respect to corrections of type (ii), it does secure exactness with respect to corrections of type (i). Through a combination of holomorphy and non-renormalization theorems typical of supersymmetric theories, supersymmetry fixes the form of the low-energy effective action of the Seiberg-Witten theory, such that the analytic expressions that Seiberg and Witten derived indeed include all the quantum corrections of the first type.\footnote{After Seiberg and Witten's original papers, there has been much work addressing the {\it high-energy regime} of this theory, i.e.~including corrections of type (ii). In short: UV extensions of ${\cal N}=2$ SYM can be given by embedding the theory in string theory or M-theory, where ${\cal N}=2$ SYM is realized as e.g.~the world-volume theory of a configuration of D-branes. The constructions involve D4-branes and NS 5-branes in Type IIA string theory or, alternatively, an M5-brane in M-theory. See, for example, \citet[pp.~459-460]{witten1997solutions}, \citet[pp.~1020-1021]{giveon1999brane}, \citet[pp.~175, 178-181]{de2004off} and \citet[pp.~239, 243, 255, 269]{gaiotto2013wall}.\label{UVlimit}} 
And, since the corrections of type (ii) are irrelevant at low energies, (i) suffices to give an exact description of the low-energy regime.\footnote{Except for some special values of the Higgs field, where the low-energy description is invalid.} By thus rendering computations more tractable and by providing results that are exact in the sense just discussed, the via media gives the Seiberg-Witten theory a higher degree of simplicity than more realistic quantum field theories: thus making it a valuable toy theory for the qualitative study of phenomena (such as the Higgs mechanism and confinement) that philosophers are interested in.

We will illustrate
the value of this via media for interpreting ontologies of quantum field theories by arguing that the Seiberg-Witten theory exhibits ontological emergence between the low-energy theory and the high-energy theory. In addition, we discuss a case of emergence in the Seiberg-Witten theory between one low-energy model and another low-energy model. We will use the notion of ontological emergence in the framework of \cite{deharo2019towards}, which characterises ontological emergence in terms of novel reference. 

The plan of the paper is as follows. In Section 2, we briefly introduce supersymmetry, its holomorphy properties, and a class of vacuum states that is identified by the supersymmetry algebra. Because these states can be individuated for a range of both weak and strong coupling, they are robust in the sense that their existence and some of their key properties in the low-energy theory are independent of the details of the high-energy theory. We argue that this makes them suitable candidates for the study of the ontology of the theory. In Section 3, we discuss the Seiberg-Witten theory. We show how type (ii) corrections appear in the Wilsonian effective action and how type (i) corrections are suppressed by holomorphy. Furthermore, we discuss the role of geometric structures and topology in the Seiberg-Witten model. Finally, in Section 4, we discuss the framework for ontological emergence in more detail and argue for the two cases of emergence mentioned above. We also argue that supersymmetry plays a central role in our claims of ontological emergence.

\section{Supersymmetry}

In their discussion of symmetries in physics, \citet[pp.~11-12]{Brading2002-BRASIP} identify four roles that symmetries play in physics and in physical theories: namely, classificatory, normative, unifying and explanatory. Our focus in this Section is on the ``normative role'' of supersymmetry in physical theories. Similar to other symmetries in quantum field theory, supersymmetry plays a ``normative role'' by imposing restrictions on the form of the action that describes a physical system: namely, by requiring that the action is invariant under a supersymmetry transformation. This limits the number of terms that can appear, thereby simplifying the calculations that are required to describe physical properties or phenomena.

In Section \ref{introS}, we provide a brief introduction to supersymmetry, superspace and superfields. Section \ref{HandNR} will consider the ``normative role'' of supersymmetry in more detail by discussing the property of holomorphy. In Section \ref{CCandBPS}, we identify a class of states in supersymmetric quantum field theories that are characterised by robustness. We will return to this class of states in Section \ref{LET}, where we will argue that their robustness makes them particularly suitable for investigating the theory's ontology.

\subsection{Introduction to Supersymmetry}\label{introS}

Supersymmetry is a spacetime symmetry that extends the Poincar\'{e} spacetime symmetries such that each boson has a supersymmetric fermion partner and vice versa. The supersymmetry generators $Q$ relate particles with different spin, namely particles with integer spin to particles with half-integer spin:
\begin{equation}
    Q\ket{\textrm{boson}} = \ket{\textrm{fermion}},~~~ Q\ket{\textrm{fermion}}=\ket{\textrm{boson}}.
\end{equation}  
In a theory with one supersymmetry, each boson is related to one fermion. In theories with \textit{extended} supersymmetry (i.e.~more than one type of supersymmetry generator $Q$), the set of states is invariant under more than one type of supersymmetry transformation, so that each boson is related to more than one fermion. Thus the number ${\cal N}$ of copies (i.e.~types) of supersymmetries, i.e.~the number of ways there are to exchange a boson and a fermion without changing the theory, is the number of independent supersymmetry generators, $Q^{I}$, where $I=1,\ldots, \mathcal{N}$. The larger the number of copies $\mathcal{N}$, the more constrained the theory is. In the Seiberg-Witten theory, where $\mathcal{N}=2$, each boson is related to two independent fermions.

Because of supersymmetry, particles in supersymmetric theories come in {\it supermultiplets}, i.e.~sets of particle states that are related to each other by the supersymmetry transformations. One way of writing down these supermultiplets is in terms of \textit{superfields}, that contain both bosonic and fermionic fields, which are defined in a generalized spacetime manifold called \textit{superspace}.\footnote{For a discussion of foundational questions on superspace, see \cite{baker2022interpreting} and \cite{Menon_superspace}.} In superspace, Minkowskian spacetime is extended to include anticommuting coordinates known as Grassmann coordinates. These coordinates allow for the inclusion of fermionic degrees of freedom in the spacetime.

In general, an extended $\mathcal{N}=2$ supersymmetric theory like the Seiberg-Witten theory can contain a hypermultiplet, which is a multiplet that includes two complex scalar fields and two fermionic fields. This hypermultiplet is taken to be either a massless state or a short BPS state (to be discussed in Section \ref{CCandBPS}), and an $\mathcal{N}=2$ massless vector multiplet.\footnote{To be more precise, an extended $\mathcal{N}=2$ supersymmetric theory in which the highest helicity of the particles in a supermultiplet is restricted to be one.} The latter contains the complex scalar field $\phi$ and its supersymmetric partner $\psi$, the gauge field $A_\mu$ and its supersymmetry partner $\lambda_\alpha$, and four states of the Dirac fermion. 

\subsection{Holomorphy and Non-renormalization} \label{HandNR}

As an extension of the Poincaré symmetries, supersymmetry can be understood as an additional constraint on the field content of a field theory. The restrictions on the theory become apparent due to the  holomorphic properties of supersymmetric theories of this kind, i.e.~the superpotential is a holomorphic function of the chiral superfields. This superpotential is a function of the superfields and coupling constants, from which the physical potential is derived. The physical potential is itself {\it not} holomorphic, but it is constrained by the holomorphic property of the superpotential. This holomorphy allows the superpotential to depend only on the chiral superfields and not on their complex conjugates, and therefore the higher-order terms in the expansion of the superfield do not appear in the superpotential. In other words, the superpotential does not receive higher-order perturbative corrections under renormalization. This is the sense in which, in the Introduction, we said that supersymmetry constrains the appearance of corrections of type (i), i.e.~corrections in the expectation value of the scalar field that renormalize the couplings in the Lagrangian.

The strength of the ``normative'' role of supersymmetry in quantum field theories lies in the fact that the restrictions it imposes often allow us to understand the theory analytically. Thus supersymmetry provides a model that is mathematically simpler than many non-supersymmetric models, thereby aiding in the interpretation of the ontology of interacting quantum field theories. We will show how supersymmetry constrains the action in the Seiberg-Witten theory in Section \ref{WEA}.

\subsection{BPS States}\label{CCandBPS}

Theories with extended supersymmetry like the Seiberg-Witten theory often contain a class of states that satisfy a lower bound on the energy, and that can be derived from the supersymmetry algebra. Our focus will be on this class of states because, as we will argue in Section \ref{emergence}, these states are encoded in the properties of the Seiberg-Witten theory that are relevant to our claim about ontological emergence. 

Theories with an extended supersymmetry can contain massive multiplets of different lengths (i.e.~with different numbers of states). The supersymmetric particle states that are described by a short massive hypermultiplet satisfy a lower energy (or mass) bound called the Bogomol'nyi–Prasad–Sommerfield (BPS) bound,\footnote{See \cite{prasadsommerfield1975, Bogomolny_eq}.}\footnote{For the time being, we use the word `particle' in the usual sense of a state with definite quantum numbers, localized in a region of spacetime. In Section \ref{SWth}, we will discuss that particle states can be limited to a certain region of parameter space, and consequently can have a limited domain of application.} which is given by: 
\begin{equation} \label{eq:BPS_mass}
    M\geq  \sqrt{2}  |Z|\,.
\end{equation} 
Here, $Z$ is given by the algebra of the generators, which for an extended supersymmetry takes the following general form:
\begin{equation}
    \left \{ Q,Q \right \} = Z\,,
\end{equation} 
where $Z$ are the central elements of the algebra, called {\it central charges}.\footnote{See \cite{Haag1975}.} 

The states that saturate the BPS bound in Eq.~(\ref{eq:BPS_mass}), i.e.~those with the lowest possible energy (or mass) allowed by the theory, are known as \textit{BPS states} or as states that are \textit{BPS saturated}. 

Because the BPS bound follows from the supersymmetry algebra, which is valid for any value of the coupling, it is invariant under renormalization, i.e.~it does not get any corrections from renormalization effects. As we discussed above, symmetry arguments suppress quantum corrections (i.e.~type (i) corrections) to the masses or charges of the BPS states. This protection against quantum corrections secures that the BPS states form a class of states that is robust, in the sense that their existence and their mass and charge are independent of the value of the coupling.

BPS states share an intimate connection with the topological and geometric aspects of a supersymmetric theory. Some of the BPS states arise in supersymmetric theories as topological soliton states, whose properties depend on the geometry of the theory. Soliton states are stable, particle-like states that are solutions to the classical, non-linear equations of motion. As such, perturbative approximations are neither valid nor sufficient to describe them in quantum field theory, and other methods are required in order to take these non-perturbative contributions into account in calculations.

Among the BPS states in the Seiberg-Witten theory are two types of topological soliton states, namely monopoles and dyons. In extended supersymmetric quantum field theories, topological solitons have topological quantum numbers that play the role of central charges.\footnote{See \cite{witten1978supersymmetry}.} In this case, $Z$ is the central charge that arises in the supersymmetry algebra in the semi-classical regime, and it is given by:
\begin{equation}\label{eq:BPS_centralcharge}
    Z= an_e+a_D n_m\, ,  
\end{equation} where $n_e$ and $n_m$ are the electric and magnetic quantum numbers of the state under consideration. Monopoles carry magnetic charge, while dyons carry both magnetic and electric charge. In the next Section, we will discuss how geometric methods such as studying monodromies are applied to find the geometric properties of the Seiberg-Witten theory, including the topological soliton states.

\section{The Seiberg-Witten Theory}\label{SWth}

The Seiberg-Witten theory is an exact low-energy effective theory of the four-dimensional $\mathcal{N}=2$ supersymmetric Yang-Mills (SYM) theory with SU(2) gauge symmetry. In 1994, Nathan Seiberg and Edward Witten developed this theory in two papers that were highly influential for the understanding of non-perturbative aspects of four-dimensional quantum field theories.\footnote{ \cite{seibergWitten1994_Confinement, Seiberg_1994_symbreaking}} Section \ref{WEA} introduces the theory and discusses the notion of the Wilsonian effective action. It shows how supersymmetry imposes restrictions on the form of the action. It also argues that the set of vacuum states of this theory constitute a moduli space, with its coordinates being the gauge-invariant expectation values of the Higgs field. In Section \ref{LET}, we continue to explore the role of geometry in Seiberg and Witten's achievement of writing down the exact low-energy effective action. We also discuss the role of dualities in unifying the low-energy models in one full theory. 

\subsection{Wilsonian Effective Action}\label{WEA}

The low-energy effective theory that was derived by Seiberg and Witten is defined by a Wilsonian effective action that is derived from the exact path integral. This method, of deriving the Wilsonian effective action, entails that the field modes with momentum higher than a certain momentum cutoff are {\it integrated out}: their effects are then included in the renormalization of the coupling constants in the Wilsonian effective action. Schematically, the general form of the action is as follows:
\begin{equation} \label{eq:action_DE}
    S = -\frac{1}{2}
   \int \dd^4x\left(g_{ij}(\phi)\,|\partial_\mu\phi^i|^2+V(\phi) + C_{n}(\phi, g, \partial g, \Lambda) (\partial \phi)^{n} \right), 
\end{equation} 
where $\phi^i$ collectively denotes the set of low-energy fields (in the low-energy model that we will consider, this set will include the complex-valued Higgs field, a gauge field, and their supersymmetric partners). The first term in this action contains the metric $g_{ij}(\phi)$ in the space of fields (which, for the Seiberg-Witten theory, only depends on the Higgs field). The second term is a real-valued potential for the scalar fields. The third term indicates schematically the high-energy corrections, i.e.~corrections of the type (ii), which we discussed in the Introduction. These are indicated schematically by operators that are higher powers in the derivatives of the fields (namely, $n>2$), which are of mass dimension higher than $4$. Here, $C_{n}(\phi, g, \partial g, \Lambda)$ is a function of the (non-renormalizable) high-energy corrections that depend on the Wilsonian momentum-cutoff $\Lambda$.\footnote{For a scalar field, the mass dimension of $(\partial\phi)^n$ is $2n$, and so since, in units where $\hbar=1$, the action is dimensionless, the mass dimension of the coefficient $C_n$ is $4-2n$ (with $n>2$). This means that, to lowest order, and neglecting curvature effects, this coefficient is proportional to $1/\Lambda^{2n-4}$. For a discussion of the significance of these higher-order terms and their dimensions, see \citet[pp.~50-54]{Williams}.} As one can see from their dimensions, these corrections are negligible at low energies, i.e.~$E\ll\Lambda$, while their contribution grows as the energy increases (namely, the coefficients are proportional to negative powers of $\Lambda$). As we will discuss later, neglecting the high-energy terms allows for the identification of a metric that describes the space of vacuum states. 

To write down the Wilsonian effective action explicitly, i.e.~to determine the forms of the metric and the potential in Eq.~(\ref{eq:action_DE}), supersymmetry arguments are used, in addition to internal symmetry arguments. This is best done using the superspace formalism that we discussed in Section \ref{HandNR}: ${\cal N}=2$ supersymmetry requires that the low-energy effective action, Eq.~(\ref{eq:action_DE}), is recovered as an integral over both spacetime and superspace, of the following form: 
\begin{equation} \label{eq:S_eff}
    S_{\sm{eff}} = \frac{1}{16\pi} \,\textrm{Im} \int \dd^4 x\, \dd^2 \theta\, \dd^2 \bar{\theta} ~\mathcal{F}(\Psi)\,, 
\end{equation} 
where $\mathcal{F}$ is the {\it prepotential} of the low-energy theory and $\theta$ and $\bar{\theta}$ are coordinates in superspace. The prepotential includes all the information about the physics in the low-energy regime, and its functional form can in principle be determined from the microscopic theory. The invariance of the action under the extended $\mathcal{N}=2$ supersymmetry is secured by the fact that the prepotential is a holomorphic quantity, both in the chiral superfield $\Psi$ and in the coupling constants, $g$. In other words, $\mathcal{F}$ only depends on the superfields and coupling constants, and not on their complex conjugates, so that $\partial \mathcal{F}(\Psi)/\partial \bar{\Psi}=0$ and $\partial \mathcal{F}(g)/\partial \bar{g}=0$. It now becomes clear how supersymmetry constrains the appearance of type (i) corrections: because higher-order terms in the perturbative expansion of the operators in the prepotential depend in some way on $\bar{\Psi}$ or $\bar{g}$, only the first-order term appears in the prepotential. Seiberg and Witten's achievement was that they were able to find the full expression for the low-energy effective prepotential, including the non-perturbative corrections.

The microscopic theory that the effective action in Eq.~(\ref{eq:S_eff}) is derived from has an SU(2) symmetry that in the low-energy regime is broken, by the Higgs mechanism of the scalar field $\phi$, down to a U(1) symmetry.\footnote{On a widespread interpretation, the Higgs mechanism does not break the local gauge symmetry, which is always present in the full theory. (For discussions, see \cite{Berghofer2023} and \cite{Maas2019}.) More precisely: only a global subgroup of the gauge group gets broken. We thank Silvester Borsboom for a discussion of this point. The model with the broken symmetry is obtained, in the low-energy regime, by expanding the bottom theory about a minimum of the potential. Thus only a U(1) symmetry remains.\label{symBfn}} 
The corresponding Higgs potential is given by:
\begin{equation}
    V(\phi)=\frac{1}{2}\, \mbox{Tr}[\phi^{\dagger},\phi]^2\geq  0\,.
\end{equation} Unbroken supersymmetry requires that the ground state satisfies $V(\phi_0)=0$. This means that either $\phi_{0}=0$, or $\phi_0^\dagger$ and $\phi_0$ need to commute. The latter case results in a family of vacuum states with $\phi_0=\frac{1}{2}\,a\,\sigma_3$, where $a$ is the vacuum expectation value of the Higgs field.

One of the consequences of the supersymmetry constraints is that the Wilsonian action describes the low-energy physics by its classical couplings. In the case of unbroken supersymmetry, these couplings are represented by parameters that can be understood as (the lowest component of) the vacuum expectation value of a superfield.\footnote{For more details on this point, see \cite{bertolini2024lectures}, chapter 9.4.} In this case, the Higgs expectation value $a$ represents the complex order parameter $u$ that corresponds to the (classical) coupling constant. The space of vacuum solutions is then parametrized by:
\begin{equation}\label{eq:Hvev}
    u = \langle\textrm{Tr}\, \phi^2\rangle = 2a^2. 
\end{equation} 
As we will discuss below, $u$ can be understood as the value of a coordinate on the manifold of vacua called the {\it moduli space}. Expectation values like $a$ are values of {\it moduli}, i.e.~a distinguished set of quantities (here, the Higgs field), that distinguish between different low-energy phases.

In the low-energy regime of the quantum theory, the moduli space is a manifold equipped with a metric that can be found by integrating over the Grassmann coordinates $\bar{\theta}$ and $\theta$ in the effective action, Eq.~(\ref{eq:S_eff}). The resulting kinetic term is then compared with the first term in Eq.~(\ref{eq:action_DE}). The overall multiplying coefficient is:
\begin{equation} \label{eq:tmetric}
    \textrm{Im}\, \tau(a):= \textrm{Im}\, \frac{\partial ^2 \mathcal{F}(a)}{\partial a^2 }\, ,
\end{equation} 
where Im$\,\tau(a)$ depends on the Higgs expectation value $a$. This coefficient can be interpreted as an overall (scale) factor of the metric on the moduli space, so that the metric is given by: 
\begin{equation} \label{eq:SW_metric_QM}
    \dd s^2 =\textrm{Im} \,\tau(a) \,\textrm{d}a\, \textrm{d}\bar{a} =\textrm{Im}\, \frac{\partial ^2 \mathcal{F}(a)}{\partial a^2 } \,\textrm{d}a\, \textrm{d}\bar{a}\, .
\end{equation} 
Since the prepotential $\mathcal{F}$ is holomorphic in $\Psi$, the coupling matrix $\tau$ only depends on the coordinate $a$, and not on its complex conjugate $\bar{a}$. The fact that it is possible to interpret this coupling matrix as a scale factor of the metric will be important in our discussion of emergence in Section \ref{Emergence}. 

In the previous discussion, we analysed the low-energy regime of the $\mathcal{N}=2$ SYM theory with SU(2) gauge symmetry: at {\it high} energies, the action given in Eq.~(\ref{eq:S_eff}) includes the additional terms, $C_{n}$. Because of these terms, the integration over $\bar{\theta}$ and $\theta$ does not allow for the extraction of an overall multiplying coefficient. Therefore, a straightforward interpretation of the {\it metric} in moduli space is lost.\footnote{The first few terms may have a geometric interpretation as corrections to this moduli space. But at high energies, one expects that this simple and neat description of the moduli space breaks down: see footnote \ref{UVlimit} for relevant work on the high-energy regime.}

\subsection{Low-Energy Models}\label{LET}

Another important property of the metric on the moduli space, given in Eq.~(\ref{eq:SW_metric_QM}), is that it is not globally defined. The facts that (i) the prepotential $\mathcal{F}(a)$ is a holomorphic (and thus harmonic) function of $a$, and (ii) its second derivative, $\mbox{Im}\,\tau(a)=\mbox{Im}\,\partial^{2}\mathcal{F}/\partial a^{2}$, is identified as a {\it running} gauge coupling (so that it is not a constant), together imply that $\mathcal{F}(a)$ cannot be positive-definite everywhere on the manifold.\footnote{In mathematical terms, the statement is that a non-constant harmonic function cannot have a minimum or a maximum within its region of definition. See Theorem 21 in \citet[p.~166]{ahlfors1979complex}.} 
But since, by (ii), $\mbox{Im}\,\tau(a)$ is the (inverse) gauge coupling $g$, a negative value of Im$\,\tau(a)$ corresponds to a negative gauge coupling, which is not a physical result. Thus there are no coordinates $a$ and $\bar{a}$ that can define the metric globally; the metric is not globally defined.

In this Section, we discuss how this problem is solved by introducing additional models that each have their own set of local coordinates. Thus by combining geometric methods and dualities, one can describe the complete moduli space of the low-energy effective theory.

Seiberg and Witten took a geometric approach to calculate the exact effective action of the low-energy effective models, namely by studying monodromies: the transformation that a mathematical object, in this case the pair ($a_D,a$), undergoes as it moves on a closed path around a singularity. Studying monodromies provides insights into the structure of the moduli space: it allows one to investigate its local and global properties, topology, algebraic geometry, and the relationships between different regions of the space. The moduli space of the Seiberg-Witten theory is a Riemann surface and, as we argued in Section \ref{WEA}, it is endowed with a metric. An investigation of the monodromies of the moduli space led Seiberg and Witten to conclude that the Riemann surface has a non-trivial topology: there are at least three singularities, at $u=\infty$, $u=\Lambda^2$ and $u=-\Lambda^2$, that are not part of the manifold itself.\footnote{The presence of two singularities additional to the singularity at $u=\infty$ agrees with the fact that the $\mathcal{N}=1$ theory, to which the $\mathcal{N}=2$ supersymmetric theory breaks upon the introduction of a mass term, possesses exactly two vacua. For more details, see \citet[ch. 12.3.2]{bertolini2024lectures}.} By studying the monodromies around each of the singularities, the prepotentials of the three low-energy models can be found.

We first focus on the region in the vicinity of the singularity $a=\infty$, i.e.~$u=\infty$, where the expectation value of the Higgs is large. Therefore, one expects usual perturbation theory to be a good approximation. Seiberg and Witten found the full expression for the prepotential in this region by expanding it in terms of the (large) expectation value of the Higgs field. They found the following expression:
\begin{equation}\label{eq:prepot}
    \mathcal{F}(a) = \mathcal{F}_{0}(a) + \frac{i}{\pi}\, a^{2}\,\textrm{log} \frac{a^{2}}{\Lambda^{2}} + \frac{a^{2}}{2 \pi i}\sum_{k=1}^{\infty} c_{k} \left ( \frac{\Lambda}{a} \right )^{4k},
\end{equation} where the first term is:
\begin{equation}
    \mathcal{F}_{0} = \frac{1}{2}\, \tau_{0} \,a^{2}\,.
\end{equation} 
This term is the classical contribution to the prepotential: which, when substituted into Eq.~(\ref{eq:S_eff}), gives the classical action, i.e.~Eq.(\ref{eq:action_DE}). The logaritmic term is the one-loop correction. The third term is the sum of non-perturbative contributions, such as the soliton contributions. In the region of the moduli space near $a=\infty$, this sum is dominated by the first two terms, i.e. the theory is dominated by the semi-classical, one-loop physics: and, as such, it is asymptotically free. For regions of the moduli space where the parameter $a$ is comparatively smaller, the sum still converges with a finite radius of convergence. In this larger region, the non-perturbative contributions cannot be neglected, and writing down the complete action requires knowing the values of the coefficients, $c_k$, of the non-perturbative terms in the sum. It is these terms that Seiberg and Witten were able to compute exactly. Using Eq.~(\ref{eq:SW_metric_QM}), one can determine the metric of this region from the prepotential in Eq.~(\ref{eq:prepot}).

Due to the limited radius of convergence of the expansion of the prepotential, this model, which we dub $M$, does not suffice to construct the complete low-energy theory. We also require models that cover the other regions of the moduli space. Thus the full theory consists of three low-energy models, $M$, $M'$ and $M''$, each of which is valid in an open region near one of the singularities. In each region, the prepotential has a different form, and it has a parameter of expansion that renders the sum convergent. As we just discussed, $M$ is valid in the region around $u=\infty$, i.e.~for large expectation values of the Higgs field. Likewise, $M'$ is valid in the region around $u=\Lambda^2$, and $M''$ is valid in the region around $u=-\Lambda^2$. These three regions of validity overlap, and together they cover the whole Riemann surface. Once the metric on the moduli space in the region around $u=\infty$ is known, one can find the appropriate coordinates for the regions described by $M'$ and $M''$ by making use of an (effective) duality transformation. In the Seiberg-Witten theory, a duality transformation allows one to shift between different coordinate patches that ascribe different local properties to the moduli space, thus shifting between different low-energy models. Because the three local models each use dual variables to describe a region of the moduli space, the prepotential and the metric are rendered non-singular by the duality transformation in the overlap region. 

This is an example of what \cite{DeHaroButterfield2024} call the {\it geometric view of theories}, i.e.~a conception of a physical theory as a geometric structure, often a differentiable manifold, and of models that are open sets that together cover the whole manifold, with (quasi-)dualities on the overlaps between the open sets. Each region of the moduli space is characterized by the vacuum expectation value of a field that plays the role of an order parameter: in our model $M$ above, this value was denoted by $a$.

Dualities can be understood as a class of symmetries, where rather than acting on \textit{solutions} to theories (as is the case with standard (gauge) symmetries), dualities act on a space of \textit{theories} (which, following \cite{deHaroButterfield2017}, we have here called `models').\footnote{As \cite{CastellaniRickles_Intro} point out, a duality transformation is more radical than an ordinary symmetry transformation. This is because, in addition to leading to a change in theoretical description, it can also lead to a change in the interpretation of the physical system. See \cite{dieksdongenharo2015} and \cite{vergouwen2022emergent} for explications of this idea in the context of emergence. The contrast between internal and external interpretations is further developed in \cite{deharobutterfield2021}.
} 
A formal definition of a duality is provided by \citet{deHaroButterfield2017}, who define a duality as an isomorphism between models that share a common core. According to this definition, a duality is a structure-preserving bijective map between the states and quantities of the two models, that is equivariant for (i.e.~that respects) their dynamics, and reflects their formal equivalence.

The dualities between the low-energy models in the Seiberg-Witten theory are in fact {\it quasi-dualities}, i.e.~they are close to dualities but they are not quite dualities: here, the duality map is only partially isomorphic, and it is not defined on the whole moduli space. Nevertheless, there is a map between the relevant state spaces that preserves the relevant subset of the quantities. This map is defined only where the models overlap on the moduli space. It is in fact an analytic continuation to a region where the series expansion of the prepotential does not converge, but can be resummed using a quasi-dual variable. 

\begin{figure}
    \centering
\tikzset{every picture/.style={line width=0.75pt}} 
\begin{tikzpicture}[x=0.75pt,y=0.75pt,yscale=-0.9,xscale=0.9]
\draw    (175.2,103.89) -- (288.66,229.96) ;
\draw [shift={(173.86,102.4)}, rotate = 48.01] [color={rgb, 255:red, 0; green, 0; blue, 0 }  ][line width=0.75]    (10.93,-4.9) .. controls (6.95,-2.3) and (3.31,-0.67) .. (0,0) .. controls (3.31,0.67) and (6.95,2.3) .. (10.93,4.9)   ;
\draw    (314.17,104.4) -- (314.17,229.96) ;
\draw [shift={(314.17,102.4)}, rotate = 90] [color={rgb, 255:red, 0; green, 0; blue, 0 }  ][line width=0.75]    (10.93,-4.9) .. controls (6.95,-2.3) and (3.31,-0.67) .. (0,0) .. controls (3.31,0.67) and (6.95,2.3) .. (10.93,4.9)   ;
\draw    (453.15,103.89) -- (339.69,229.96) ;
\draw [shift={(454.49,102.4)}, rotate = 131.99] [color={rgb, 255:red, 0; green, 0; blue, 0 }  ][line width=0.75]    (10.93,-4.9) .. controls (6.95,-2.3) and (3.31,-0.67) .. (0,0) .. controls (3.31,0.67) and (6.95,2.3) .. (10.93,4.9)   ;
\draw    (202.65,75.61) -- (289.21,75.61) ;
\draw [shift={(291.21,75.61)}, rotate = 180] [fill={rgb, 255:red, 0; green, 0; blue, 0 }  ][line width=0.08]  [draw opacity=0] (12,-3) -- (0,0) -- (12,3) -- cycle    ;
\draw [shift={(200.65,75.61)}, rotate = 0] [fill={rgb, 255:red, 0; green, 0; blue, 0 }  ][line width=0.08]  [draw opacity=0] (12,-3) -- (0,0) -- (12,3) -- cycle    ;
\draw    (342.96,75.61) -- (428.25,75.61) ;
\draw [shift={(430.25,75.61)}, rotate = 180] [fill={rgb, 255:red, 0; green, 0; blue, 0 }  ][line width=0.08]  [draw opacity=0] (12,-3) -- (0,0) -- (12,3) -- cycle    ;
\draw [shift={(340.96,75.61)}, rotate = 0] [fill={rgb, 255:red, 0; green, 0; blue, 0 }  ][line width=0.08]  [draw opacity=0] (12,-3) -- (0,0) -- (12,3) -- cycle    ;
\draw (308.17,247.87) node [anchor=north west][inner sep=0.75pt]    {$T$};
\draw (302,66.74) node [anchor=north west][inner sep=0.75pt]    {$M'$};
\draw (159.76,66.74) node [anchor=north west][inner sep=0.75pt]    {$M$};
\draw (448,66.74) node [anchor=north west][inner sep=0.75pt]    {$M''$};
\draw (200.36,47.33) node [anchor=north west][inner sep=0.75pt]  [font=\normalsize]  {$\mbox{S-duality}$};
\draw (338.57,47.33) node [anchor=north west][inner sep=0.75pt]  [font=\normalsize]  {$\mbox{T-duality}$};
\end{tikzpicture}
    \caption{\small Overview of the Seiberg-Witten theory. The vertical arrows indicate taking the low-energy regime of the microscopic theory $T$, which results in three effective models, $M$, $M'$, and $M''$. These models are related to each other by quasi-dualities, indicated by the horizontal arrows.}
    \label{fig:Overview}
\end{figure}
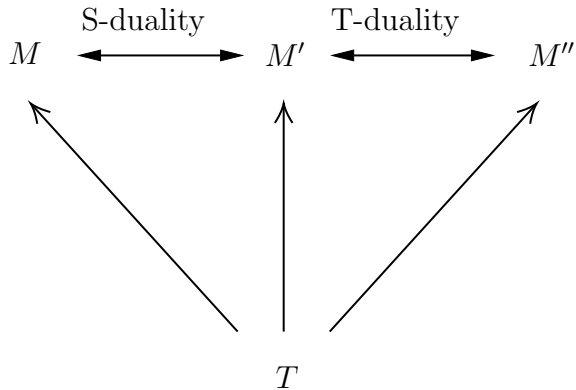

The models $M$ and $M'$ are related to each other by S-duality: see Figure \ref{fig:Overview}. S-duality transformations are also known as `weak-strong coupling dualities', because they relate the physics in a strong-coupling regime, where perturbative methods break down, to the physics in a weak-coupling regime, where perturbation theory gives a valid approximation of the physical system. More specifically, this S-duality is a quasi-duality of the electromagnetic type, which means that the models $M$ and $M'$ are quasi-duals under the exchange of an {\it  electrically charged gauge boson and a magnetically charged monopole}. Thus it relates a weak-coupling gauge theory to a strong coupling theory of solitons, i.e.~topological particles.\footnote{For a fuller account of these aspects of the Seiberg-Witten theory, and in particular of the philosophical significance of solitons, see \citet{DeHaroButterfield2024}.}

The models $M'$ and $M''$ are related to each other by a T-duality transformation (not to be confused with T-duality in string theory), which typically results in a constant overall shift in the action. The $S$- and $T$-dualities of the Seiberg-Witten theory are the two 
generators of an $\mbox{SL}(2,\mathbb{Z})$ symmetry group, which relates different points on the moduli space, and so this is the symmetry group of the moduli space.

A physical interpretation of the singularities in the moduli space is to understand them as massive particles, either an elementary particle or a soliton, that have been integrated out.\footnote{See 't \cite{hooft1981topology}, pp.~463-465.} At these values of $a$, the particles become massless and the low-energy approximation is invalid. The duality transformations allow one to set up an effective action that describes the regions around the singularities, where the mass of these particles is small and therefore the particles need to be included in the effective action. 

Two of the singularities, at $u=\Lambda^2$ and $u=-\Lambda^2$, can be interpreted as corresponding to massless solitons, respectively a monopole and a dyon. The monopole solutions in this theory were initially described in 1974 by Gerard \citeauthor{tHooft1974magneticmonopoles} and Alexander Markovich \citeauthor{Polyakov:1974}. In examining the Georgi-Glashow model, they observed that monopoles occur naturally in Yang-Mills theories with a Higgs mechanism. These solutions correspond to a conserved magnetic current that is topological in nature. The third singularity, at $u=\infty$, can be interpreted as corresponding to a massless gauge boson. The soliton solutions are dual to each other, and to the elementary particle solutions. Performing a full $\mbox{SL}(2,\mathbb{Z})$ duality transformation on the gauge particle state corresponding to the singularity at $u=\infty$ gives a dyonic state that corresponds to the singularity at $u=-\Lambda^2$. These three particles, the monopole, the dyon, and the gauge boson, are part of the BPS spectrum of the theory.

\section{Emergence in the Seiberg-Witten Theory}\label{Emergence}
 
In this Section, we will argue that the relation between the ${\cal N}=2$ supersymmetric SU(2) Yang-Mills theory and the low-energy Seiberg-Witten theory, that we discussed in the previous Section, is best conceptualised as a relation of emergence. Namely, there are two main benefits from bringing in a conception of emergence: First, Section \ref{emergence} will illustrate how Butterfield's conception of emergence, as `properties or behaviour of a system which are {\it novel} and {\it robust} relative to some appropriate comparison class' (Butterfield, 2011:~p.~921), casts light on the physical interpretation of the Seiberg-Witten theory. Thus Section \ref{emergence} will use the idea of a `delicate balance' between reduction and emergence, that has been recently advocated by several authors, to conceptualise the relations between the various models involved in the Seiberg-Witten theory.\footnote{See e.g.~Butterfield (2011), Guay and Sartenaer (2016), Crowther (2016) and De Haro (2019).} 
Second, Section \ref{EGP} will spell out which geometric properties emerge in the low-energy regime. Thus Section \ref{EGP} will thereby illustrate, in a concrete example, how a theory of emergence can benefit from precise case studies in quantum field theory such as the Seiberg-Witten theory. For, in so far as Wilsonian renormalization group methods are very general and can be applied widely in quantum field theory, the Seiberg-Witten theory is paradigmatic of the relations between high-energy and low-energy theories, and it casts light on the role of topology in a theory of emergence. 

\subsection{Framework for Emergence}\label{emergence}

We will use as our framework for emergence a theory of ontological emergence developed in De Haro (2019). The leading idea is that ontological emergence is a matter of novelty in the semantics of a physical theory, relative to an appropriate comparison class. (This is best explained in the most general case, where the comparison class is taken to be another theory or set of theories: as against the special case where the comparison class is the same theory, and only some properties or behaviours are different.)\footnote{See Butterfield (2011) at p.~921.} 
Thus we distinguish a bottom theory, $T_{\sm b}$, and a top theory or model, $T_{\sm t}$ or $M_{\sm t}$. In our case, it will initially be appropriate to have a model at the top, because as we will discuss, there are three possible linkage relations, each with a model in its codomain. Together, the three models give what we call the Seiberg-Witten theory: see Figure \ref{fig:Overview}.

Thus the idea of emergence that we are going to discuss is, roughly, that the bottom and top theories (or models) (i) are formally linked, and (ii) their semantics ``come apart'': namely, the linkage relation does not respect the semantics, so that a change in the semantics is required. To realize this contrast between the formal linkage and the change in semantics, we use the following (standard) distinction between (i) a bare theory or model, and (ii) its interpretation in a domain of application:

(i)~~{\it The bare theories and the linkage map.} A bare i.e.~uninterpreted theory or model is (also standardly) described as a triple of state-space, set of quantities, and dynamics, which we denote by: $\langle{\cal S},{\cal Q},{\cal D}\rangle$. The set of quantities ${\cal Q}$ will be the algebra of gauge-invariant fields, which give representations of the super-Poincar\'e algebra with various quantum numbers (they are interpreted as masses and spin quantum numbers, see Section \ref{introS}). Thus taking $T$, i.e.~the ${\cal N}=2$ SU(2) SYM theory, as our bottom theory, these are the fields in the corresponding supermultiplet (viz.~a vector supermultiplet), while in the top model $M$, i.e.~${\cal N}=2$ U(1) SYM, these are the fields of the chiral multiplet.

For quantum theories, the set of states ${\cal S}$ is usually a Hilbert space, ${\cal H}$. In our bottom theory, the Hilbert space is spanned by the irreducible representations of the super-Poincar\'e algebra: and, as we discussed in Section \ref{CCandBPS}, we will focus on the subspace of BPS states, i.e.~the states that satisfy the BPS bound in Eq.~(\ref{eq:BPS_mass}), so that ${\cal S}$ will be restricted to this subset. In our top theory, the BPS states are encoded in the properties of the moduli space (especially, in the monodromies around each of the singularities), which is the space of low-energy vacua (not just for a single model, but for the whole Seiberg-Witten theory). Thus we will first take the state-space of the top model to be the open region of the moduli space corresponding to a single model: and, after we discuss the relation between the different models, we will extend the state-space to the whole moduli space, so that we will regard the whole low-energy Seiberg-Witten theory, as described by its moduli space, to be emergent.

The fields take values on the states, i.e.~for a state $\psi\in{\cal H}$, a gauge-invariant field $\Phi\in{\cal Q}$ is assigned an expectation value $\bra \psi\Phi\ket \psi\in\mathbb{C}$. This allows us to think of a field $\Phi\in{\cal Q}$ as an assignment of a complex number to a state, i.e.~$\Phi:{\cal H}\rightarrow\mathbb{C}$. Thus in the low-energy model $M$ discussed in Section \ref{SWth}, the gauge-invariant expectation value of the Higgs field, Eq.~(\ref{eq:Hvev}), is a coordinate on the space of low-energy vacua, i.e.~the moduli space. Each of the other two models has an analogous order parameter, i.e.~the expectation value of a distinguished complex-valued field that characterises a macroscopic phase of the low-energy theory, which are obtained by quasi-duality. (Since the different open sets of the moduli space have qualitatively different ``macroscopic'' properties, this is reminiscent of phase transitions in statistical mechanics: hence our use of the word `phase'.) The order parameters of the other models give different coordinatizations of the moduli space that are valid in the other open sets (although we will not need to discuss their details). 

The dynamics is given by a choice of a self-adjoint operator that is the theory's Hamiltonian (alternatively, by a choice of a Lagrangian, which in the low-energy model $M$ is given by Eq.~(\ref{eq:action_DE}), and analogously for the other models).

The {\it linkage} between the bottom and top theories that the framework for emergence requires is a non-injective, partial map, that is a partial homomorphism, i.e.~it respects part of the theory's structure. This map has as its inputs states of the bottom theory, and as its outputs states of the top theory (likewise, there is a similar map for quantities). The non-injectivity requirement embodies the idea of `coarse graining', so that a state of the top theory at low energies is the output of a (usually large) number of input states in the domain of the linkage map, i.e.~the bottom theory at high energies.

It will be useful to think of the linkage map(s) in terms of the variation of a parameter in the path integral formulation of the theory, together with the associated procedure of integrating out the modes that exceed the relevant scale: namely, the energy, in units of the cutoff, i.e.~$E/\Lambda$, so that as we vary this parameter we go through a sequence of states and quantities that are relevant at that energy (recall the discussion of Wilsonian renormalization, in Section \ref{WEA}). The image of the map(s), i.e.~the top model, is obtained when this parameter is taken to zero or a value close to zero, so that only the massless modes remain.\footnote{We say `or a value close to zero' because it is only required is that the energy $E$ is much smaller than the mass of the lightest massive mode that has been integrated out. Thus, although it is useful to think in terms of taking the limit $E/\Lambda\rightarrow 0$, nothing here hinges on actually being at the limit.} 
The comparison class is the class of theories at larger values of this parameter (in particular, very large ones). The non-injectivity requirement will be satisfied, because, regardless of the value of the parameter $E/\Lambda$ at which we take our bottom theory, and regardless of what this theory looks at high energies, we always get the same top model. Also, the map is partly structure-preserving if the bottom and top theories share part, but not all, of their (super)symmetries: as in the case of symmetry breaking, and of bottom and top theories that share a supersymmetry algebra, which we are interested in here. Furthermore, the linkage map forgets the detailed information about the fields whose masses exceed the cut-off: namely, the top theory does not distinguish between members of the comparison class that have the same properties at low energies, but different properties at high energies. Thus the low-energy theory is insensitive to part of the detail of the bottom theory (especially, about the higher-order terms in Eq.~(\ref{eq:action_DE})), even though it contains the detailed non-perturbative information about the coupling parameter, through the non-perturbative terms in the superpotential, Eq.~(\ref{eq:prepot}) (recall our discussion of the double expansion, in Section \ref{intro}).

In the case at hand, we take the two linkage maps just discussed, between the sets of states and between the sets of quantities, to be cases of Nagelian reduction.\footnote{Recall that, in short, a theory $T_{\tn t}$ is Nagel-reduced to $T_{\tn b}$ iff $T_{\tn t}$ can be deduced from $T_{\sm b}$, using appropriate bridge laws. Here, the deduction is the derivation of the low-energy Wilsonian effective action from the exact path integral. See \cite{nagel1961}, \cite{butterfield2011less}, and \cite{dizadji2010s}.} Namely, the quantum numbers that characterize the BPS states of any of the three models at the top are subsets of the quantum numbers of the BPS states of the bottom theory (these numbers follow from the supersymmetry algebra, and in particular from Eqs.~(\ref{eq:BPS_mass}) and (\ref{eq:BPS_centralcharge})). Thus, in effect, the linkage map is a {\it bridge law} between the states of the bottom and top theories.\footnote{Bridge laws are used here because the state-space of the top theory is not simply a subset of the state-space of the bottom theory. The bottom and top states have different properties: for example, they are not in the same representation, because they appear in different supersymmetry multiplets. Thus one needs to specify how the states of the bottom theory are mapped onto those of the bottom theory.}
Likewise, the quantities of the top model, especially the Wilsonian effective action (from which other quantities can be derived), are derived from those of the bottom theory, through the renormalization procedure and symmetry arguments (themselves based on the supersymmetry of the bottom theory) discussed in Section \ref{WEA}.

Hence our claim, in the preamble of this Section, that the relation between the bottom and top theories involves a `delicate balance' between reduction and emergence. For we can Nagel-reduce the structure of the top models to the bottom theory (adding, where appropriate, definitions of terms and other bridge laws). And at the same time, as point (ii) and Section \ref{EGP} will argue, there is ontological emergence, i.e.~referential novelty.

Agreed: as Section \ref{EGP} will discuss, even before we interpret anything, novel structures already appear in the top theory. But this would not by itself give us ontological emergence, if those structures gave merely different formal descriptions of the same domain of application. Thus we will argue that there is referential novelty, so that the linkage and interpretation maps do not commute. We do this in (ii) below. Section \ref{EGP} will then discuss in more detail the properties that are emergent.

(ii)~~{\it Interpretation maps and a condition for not meshing.} A familiar, indeed mainstream, way to present the semantics of a physical theory is to use referential semantics. We will endorse this framework, which allows us to model the semantics by an interpretation map from states and quantities in the bare theory to a domain of application in the world. 

In particular, as we discussed above, the distinguished set of fields that are the order parameters distinguish the macroscopic i.e.~low-energy phases of a quantum field theory, which are the equivalence classes of states with the same qualitative behaviour. Thus we can think of the values of quantities, interpreted as indicators for various phases, as providing a semantics for the states. 

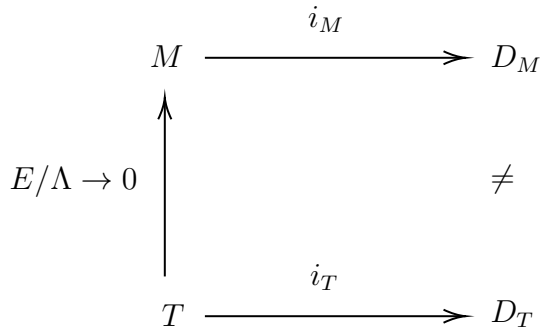
\begin{figure}
\centering
\tikzset{every picture/.style={line width=0.75pt}} 
\begin{tikzpicture}[x=0.75pt,y=0.75pt,yscale=-1,xscale=1]
\draw [line width=0.75]    (170,50) -- (298,50) ;
\draw [shift={(300,50)}, rotate = 180] [color={rgb, 255:red, 0; green, 0; blue, 0 }  ][line width=0.75]    (10.93,-3.29) .. controls (6.95,-1.4) and (3.31,-0.3) .. (0,0) .. controls (3.31,0.3) and (6.95,1.4) .. (10.93,3.29)   ;
\draw [line width=0.75]    (170,180) -- (298,180) ;
\draw [shift={(300,180)}, rotate = 180] [color={rgb, 255:red, 0; green, 0; blue, 0 }  ][line width=0.75]    (10.93,-3.29) .. controls (6.95,-1.4) and (3.31,-0.3) .. (0,0) .. controls (3.31,0.3) and (6.95,1.4) .. (10.93,3.29)   ;
\draw [line width=0.75]    (150,160) -- (150,72) ;
\draw [shift={(150,70)}, rotate = 90] [color={rgb, 255:red, 0; green, 0; blue, 0 }  ][line width=0.75]    (10.93,-3.29) .. controls (6.95,-1.4) and (3.31,-0.3) .. (0,0) .. controls (3.31,0.3) and (6.95,1.4) .. (10.93,3.29)   ;
\draw (147,172.4) node [anchor=north west][inner sep=0.75pt]    {$T$};
\draw (141,42.4) node [anchor=north west][inner sep=0.75pt]    {$M$};
\draw (221,152.4) node [anchor=north west][inner sep=0.75pt]    {$i_{T}$};
\draw (220,22.4) node [anchor=north west][inner sep=0.75pt]    {$i_{M}$};
\draw (311,42.4) node [anchor=north west][inner sep=0.75pt]    {$D_{M}$};
\draw (311,170.4) node [anchor=north west][inner sep=0.75pt]    {$D_{T}$};
\draw (311,102.4) node [anchor=north west][inner sep=0.75pt]    {$\neq $};
\draw (71,102.4) node [anchor=north west][inner sep=0.75pt]    {$E/\Lambda\rightarrow 0$};
\end{tikzpicture}
\caption{\small Ontological emergence of the low-energy model $M$ of the Seiberg-Witten theory from the microscopic theory $T$. The interpretation maps $i_{M}$ and $i_{T}$ require different semantics, hence the linkage and the interpretation fail to commute and the domain of applications $D_{M}$ and $D_{T}$ differ. }\label{fig:E_TM}
\end{figure}

With these preliminaries, ontological emergence is the phenomenon that, when we compare the high- and low-energy theories, although we can relate the states and quantities of the bottom and top bare theories formally, using the partly structure-preserving linkage map, this linkage does not mesh with the interpretation, so that the two maps form a diagram that does not commute (see Figure \ref{fig:E_TM}). Namely, as we move from the bottom to the top by taking $E/\Lambda$ to a small value, the semantics of the bottom theory stops being applicable, and we need to construct a new semantics, i.e.~a new interpretation map, for the top model.\footnote{For some examples of e.g.~how the interpretation of the bottom theory becomes inconsistent with the top theory, see \citeauthor{deharo2019towards} (2019:~p.~35-48). This means that there is no limiting system associated with that interpretation, and a new interpretation is required.} 
In particular, compared to the domain of application of the bottom theory, the domain of application of the top model contains novel properties. The appearance of these novel properties, which is signalled by the need to change the interpretation map, and hence by the non-commutativity of the diagram, is what we call {\it ontological emergence}. In the next Section, we will discuss these novel properties. 

\subsection{Novelty in the Seiberg-Witten Theory} \label{EGP}

According to the definition given above, the mark of ontological emergence is the appearance of novel properties in the domain of application of the top theory, compared to the bottom theory: such that the linkage and interpretation maps form a non-commuting diagram, as in Figure \ref{fig:E_TM}. In this Section, we will first focus on the novel properties that arise in the top theory, i.e.~the low-energy theory, and then argue that the three low-energy models $M, M'$, and $M''$ individually, as well as the complete Seiberg-Witten theory, ontologically emerge from the microscopic theory $T$, i.e.~the $\mathcal{N}=2$ super Yang-Mills theory with SU(2) gauge symmetry.  We end this Section by discussing an alternative case of emergence, namely the relative emergence between the low-energy models that make up the Seiberg-Witten theory. We also discuss the role of topology in these cases of emergence.

For simplicity, we focus on one of the three local low-energy effective field theory models and identify this as our top model. We take the model $M$ that describes the region on the moduli space around the singularity $u=\infty$. This model is an $\mathcal{N}=2$ super Yang-Mills theory with U(1) gauge symmetry. Figure \ref{fig:E_TM} gives a schematic depiction of this case of emergence.

We will discuss the emergence of this model in two steps, both of which illustrate the previous Section's statement that, as we take $E/\Lambda$ to be small or zero (namely, much smaller than the mass of the lightest massive mode that has been integrated out), `the semantics of the bottom theory stops being applicable'. Thus we first discuss novel features of the top domain that are not in the domain of the bottom theory, and then we discuss features of the bottom theory that are inconsistent with the top theory. In other words, some concepts that appear in the bottom theory stop being applicable in the top theory.\\
\\
{\it Novel features of the top domain that are not in the bottom theory's domain.} We begin with the low-energy effective action. As Section \ref{WEA} shows, the action contains non-renormalizable terms that, at sufficiently high energies, prohibit us from identifying a metric that describes a classical moduli space. However, as $E/\Lambda$ becomes small, these terms are negligible and the vacua of the theory form a manifold, in which physical quantities take on a value, and which is equipped with geometric structures such as the metric, given by Eq.~(\ref{eq:SW_metric_QM}). Thus to get from the top bare theory to the domain of application, we interpret these mathematical structures as physical properties of the theory: especially, as a metric for angles and distances between vacuum states. Since the interpretation map $i_T$ of the bottom theory does not have a metric in its range, while $i_M$ does have a metric in it range, this illustrates the non-commuting diagram in Figure \ref{fig:E_TM}: for small $E/\Lambda$, there is a novel property, i.e.~a metric on a one-dimensional complex manifold, in the theory's domain of application. In addition to being {\it novel}, the metric is also a {\it robust} property.\footnote{ We follow the notion of robustness given in \cite{butterfield2011emergence}. See also \cite{franklin2018emergence}, Section 3.2, for a discussion on robustness in the context of emergence.} The robustness we consider here is similar to the case of the emergence of the liquidity of water in the continuum limit: once one starts to zoom in on the surface of water, one begins to see the granularity and the internal structure of the molecules---and yet we say that liquidity is a robust property of water. The liquidity does not disappear because at smaller distances it begins to be granular rather than continuous.\footnote{See for example Butterfield’s discussion of ‘emergence before the limit’ in \cite{butterfield2011less}.} Likewise, in our case, the low-energy theory $M$ is insensitive to part of the detail of the high-energy theory $T$.  

Furthermore, the classical properties that arise in the domain of application of the low-energy models are novel compared to the microscopic quantum theory. For example, in the low-energy model $M$, the Higgs field behaves classically.\footnote{By `behaves classically', we do not mean that the Higgs field has a specified classical value, but rather that its correlation functions factorise, because in the low-energy regime a single configuration dominates the path integral, and other contributions are suppressed. See e.g.~\citet[Section IV]{witten19801} and \citet[p. 392]{coleman1985aspects}.} These novelties are the result of the classical and geometric nature of the moduli space, and require a new semantics as discussed above. 

Finally, the properties of the BPS multiplets are novel in the top theory's domain of application compared to the bottom theory. Even though the $\mathcal{N}=2$ supersymmetry holds both in the low-energy model and in the high-energy theory, the BPS states are in different representations of the supersymmetry algebra in the top theory compared to the bottom theory, i.e.~the supermultiplets are not fully preserved under the linkage map. Thus states that are related by the linkage map have the same mass and charge, but different spins. Thus they are different states altogether and, in addition to being novel, they are also robust. For as we discussed in Section \ref{CCandBPS}, the BPS bound is dictated by the supersymmetry algebra, and therefore the relation between the masses of the BPS states and their charges is invariant under the renormalization group, and remains valid after perturbative and non-perturbative contributions are taken into account. In other words, if the states in the microscopic theory satisfy the BPS bound, they must satisfy the bound in the macroscopic theory as well. Furthermore, these states exist at any value of the coupling. Perturbations in the underlying physics, i.e.~the inclusion of perturbative or non-perturbative terms in the action, do not affect the appearance of the BPS states in the full low-energy theory, even though they are not all present in each phase, i.e.~in each of the models, of the low-energy theory. In this sense, supersymmetry secures the robustness of the BPS states, more specifically the gauge states, the monopole states and the dyonic states, of the Seiberg-Witten theory. This robustness makes them {\it reliable parts of the ontology of the theory}: namely, such states are present in the domain of application at low energies, no matter how large the quantum corrections, i.e.~for any value of the coupling. \\  
\\
{\it Features of the bottom theory that are inconsistent with small $E/\Lambda$}. The bottom theory has an SU(2) symmetry group, with a corresponding number of gauge bosons (namely, three), only one of which survives at small $E/\Lambda$, i.e.~under the linkage map. Thus in the low-energy regime, only a U(1) subgroup of the original gauge symmetry remains.\footnote{That the {\it low-energy model} has a U(1) symmetry and no SU(2) symmetry is independent of the interpretative details of the Higgs mechanism. The model with the broken symmetry at the top is {\it different} from the bottom theory, because it is obtained by expanding the bottom theory about a minimum of the potential, and taking the low-energy limit. In this regime there is only a U(1) symmetry. See also footnote \ref{symBfn}.} 
Therefore, statements such as `the gauge field has an SU(2) gauge symmetry', which can be made and have a referent in the bottom theory, are inconsistent with the linkage map: such statements have no referent in the top model, i.e.~no corresponding item in the top domain of application. Thus again, this means that the linkage and interpretation maps do not commute.\\

The other two low-energy models similarly emerge from the microscopic theory. Each model is related to the bottom theory by a linkage map that maps the microscopic theory to the sub-region of the moduli space that is described by that low-energy model. Let us briefly discuss the emergence of $M'$ from $T$. Here, $M'$ is the $\mathcal{N}=2$ supersymmetric version of QED with a light $\mathcal{N}=2$ hypermultiplet. By including an (approximate) S-duality transformation in the linkage map, the model $M'$ emerges from the microscopic theory $T$. In this case, the monopoles obtain electric, rather than magnetic, charge upon the duality transformation. Therefore, we can say that the metric of the regions of the moduli space and the classical properties described by $M'$ emerge from the microscopic theory.

Taking into account the S- and T-duality transformations that unify the low-energy models to form a complete low-energy effective action, and the fact that the BPS relation given in Eq.~(\ref{eq:BPS_centralcharge}) is invariant under the duality, we argue that the moduli space of vacua of the whole Seiberg-Witten theory ontologically emerges from the microscopic theory.\\ 

\begin{figure}[t]
\centering
\tikzset{every picture/.style={line width=0.75pt}} 
\begin{tikzpicture}[x=0.75pt,y=0.75pt,yscale=-1,xscale=1]
\draw [line width=0.75]    (170,50) -- (298,50) ;
\draw [shift={(300,50)}, rotate = 180] [color={rgb, 255:red, 0; green, 0; blue, 0 }  ][line width=0.75]    (10.93,-3.29) .. controls (6.95,-1.4) and (3.31,-0.3) .. (0,0) .. controls (3.31,0.3) and (6.95,1.4) .. (10.93,3.29)   ;
\draw [line width=0.75]    (170,180) -- (298,180) ;
\draw [shift={(300,180)}, rotate = 180] [color={rgb, 255:red, 0; green, 0; blue, 0 }  ][line width=0.75]    (10.93,-3.29) .. controls (6.95,-1.4) and (3.31,-0.3) .. (0,0) .. controls (3.31,0.3) and (6.95,1.4) .. (10.93,3.29)   ;
\draw [line width=0.75]    (150,160) -- (150,72) ;
\draw [shift={(150,70)}, rotate = 90] [color={rgb, 255:red, 0; green, 0; blue, 0 }  ][line width=0.75]    (10.93,-3.29) .. controls (6.95,-1.4) and (3.31,-0.3) .. (0,0) .. controls (3.31,0.3) and (6.95,1.4) .. (10.93,3.29)   ;
\draw (147,172.4) node [anchor=north west][inner sep=0.75pt]    {$M$};
\draw (141,42.4) node [anchor=north west][inner sep=0.75pt]    {$M'$};
\draw (220,22.4) node [anchor=north west][inner sep=0.75pt]    {$i_{M'}$};
\draw (221,152.4) node [anchor=north west][inner sep=0.75pt]    {$i_{M}$};
\draw (311,42.4) node [anchor=north west][inner sep=0.75pt]    {$D_{M'}$};
\draw (311,170.4) node [anchor=north west][inner sep=0.75pt]    {$D_{M}$};
\draw (311,102.4) node [anchor=north west][inner sep=0.75pt]    {$\neq $};
\draw (71,102.4) node [anchor=north west][inner sep=0.75pt]    {$\mbox{Duality}$};
\end{tikzpicture}
\caption{\small Ontological emergence of the low-energy model $M'$ from the low-energy model $M$. The interpretation maps $i_{M}$ and $i_{T}$ require different semantics, hence the linkage and the interpretation fail to commute and the domain of applications $D_{M'}$ and $D_{M}$ differ.}\label{fig:E_MM}
\end{figure}
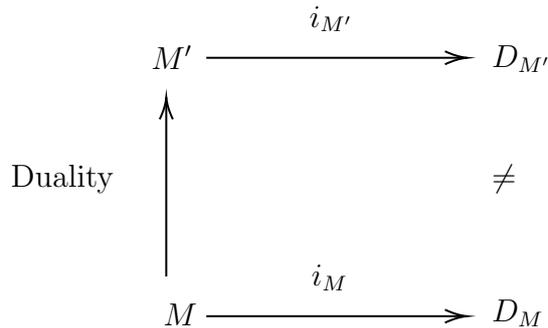

{\it Topology of the moduli space and emergence.} As an alternative to the above discussion, of looking at emergence relative to the high-energy theory, one can also look at the relative emergence between the three low-energy models.\footnote{See also \cite{vergouwen2022emergent}. For more details about the topological argument, and an analogy with a phase transition towards a ferromagnetic state, see \cite{DeHaroButterfield2024}.} 
Thus for example, one may consider the emergence, as one moves from any point in the moduli space far from $u=\Lambda^2$ towards $u=\Lambda^2$, of a monopole state. In this type of emergence between phases, the topological non-triviality of the moduli space, and specifically the fact that the moduli space cannot be covered by a single open set, plays an important role:\footnote{Although two open sets suffice to cover the punctured plane, the physical properties of the Seiberg-Witten theory (namely, the lack of convergence of the prepotential) require us to adopt three models: and thus to cover the Riemann surface with three open sets.}
for it implies that we cannot use a single model. Thus we need three models to cover the whole moduli space, hence the possibility of emergence: more specifically, each open set comes equipped with a different complex coordinate (namely, the order parameter for that model). These complex coordinates are maps from the given open set onto $\mathbb{C}$, with each open set mapping onto its own copy of $\mathbb{C}$, i.e.~its own coordinate space. These coordinate spaces are {\it model spaces} (see \cite{Lee2018}, p.~55), i.e.~spaces of zero or constant curvature that model (regions of) a Riemannian space. These model spaces provide a model-theoretic semantics for the moduli space and the quantities defined on it that guide the physical interpretation of the models. In other words, these model spaces are structures for the domains of application shown in Figure \ref{fig:E_MM}.

In sum: the topology of the moduli space implies that we require more than one model space to coordinatize the moduli space. Each model space describes a domain of application, and so the domains of application corresponding to each region are also different from each other. Thus in the low-energy theory, ontological emergence, which is characterised by the differences between the domains of application, depends on the topologically non-trivial structure of the moduli space, and is vividly illustrated by the fact that the moduli space has {\it three different model spaces}.\footnote{For a discussion of the implications of moduli space for the formulation of scientific theories, see \cite{DeHaroButterfield2024}.}

\section{Conclusion and Outlook}

The primary aim of this paper has been to illustrate that the combination of a double expansion and supersymmetry serves as a {\it via media} that assists in bridging the gap between the ontologies of an exact theory and that of its semi-classical limit. We have illustrated how the techniques developed for the Seiberg-Witten theory can be used to learn about non-perturbative quantum field theory, particularly by employing reduction and emergence: and how supersymmetry, along with the other techniques being used, is helpful in this regard. 

The two expansions that we have distinguished result in two possible types of corrections to the effective action of the Seiberg-Witten theory: (i) corrections in the expectation value of the scalar field, and (ii) corrections by higher-dimensional operators for high-energy modes that contribute at higher energies. We showed that, by constraining the number of terms in the action and ensuring exactness in type (i) corrections, supersymmetry assists in making well-founded statements about the ontology of interacting quantum field theories such as the Seiberg-Witten theory. Namely, it can be used to determine the properties of BPS states such as their masses, electric and magnetic charges, and spin quantum numbers, as well as the topological and geometric properties of the moduli space and its set of quantities, such as the order parameters and the free energy.

Thus our discussion contributes to the debate about how to answer interpretative questions in quantum field theory. This debate hinges on a problem well-known to philosophers: that of how to rigorously define quantum field theories. This latter problem is sometimes discussed in terms of a contrast between algebraic and Lagrangian quantum field theories.\footnote{See \cite{fraser2011takeseriously}, \cite{wallace2006defence,wallace2011takingseriously}, and \cite{ruetsche2002interpreting}.}
While the algebraic approach is mathematically rigorous, it has so far not been very practically applicable to interacting four-dimensional quantum field theories. And, while the Lagrangian approach {\it is} thus applicable and is empirically well-confirmed, it is in general not mathematically rigorous, which invites questions about the validity of the philosophical conclusions that one may draw from the interpretation of such theories. Following \cite{fraser2009quantum}, we characterise two aims that align with these approaches: to interpret those quantum field theories that are used in scientific practice, commonly formulated within the Lagrangian approach to quantum field theory, and to clarify foundations and provide interpretations, which is more easily achieved by focussing on more rigorous versions of QFT, such as the algebraic formulations. Through the via media that we have illustrated, we argue that we are faced with the prospect of bringing together these two aims. Namely, we have shown how supersymmetry plays a ``normative'' role in quantum field theory that aids in computing exact, rather than (semi-classically) approximate, and therefore mathematically more rigorous, solutions. These solutions are valid to all orders in the coupling parameter, and so they are fully quantum. This makes the Seiberg-Witten model particularly suited for interpretive questions in quantum field theory.\footnote{A more extensive discussion of the philosophical relevance of studying Lagrangian quantum field theories with exact results can be found in \cite{vergouwen2022emergent}.}
A complete understanding of (the theoretical content of) quantum field theory of course also requires understanding of the high-energy limit, and for that the simplest Seiberg-Witten theory that we have discussed here does not suffice.\footnote{The high-energy limit of ${\cal N}=2$ SYM that is obtained by embedding this theory in string theory or M-theory is discussed in, for example, \cite{witten1997solutions}, \cite{giveon1999brane}, \cite{de2004off} and \cite{gaiotto2013wall} (see also footnote \ref{UVlimit}).}

\subsection*{Acknowledgements}
\addcontentsline{toc}{section}{Acknowledgements}

We thank Guido Bacciagaluppi,  Silvester Borsboom, Jeremy Butterfield and Enrico Cinti for discussions. SDH thanks audiences at Oxford, Utrecht and Nijmegen, where part of this work was presented. We also thank two anonymous reviewers for their comments.

\bibliography{Bibtex}

\begin{thebibliography}{41}
\expandafter\ifx\csname natexlab\endcsname\relax\def\natexlab#1{#1}\fi
\expandafter\ifx\csname url\endcsname\relax
  \def\url#1{{\tt #1}}\fi
\expandafter\ifx\csname urlprefix\endcsname\relax\def\urlprefix{URL }\fi

\bibitem[{Ahlfors(1979)}]{ahlfors1979complex}
Ahlfors, L.~V. (1979).
\newblock Complex analysis: an introduction to the theory of analytic functions
  of one complex variable.
\newblock {\em New York, London\/}, {\em 177\/}.

\bibitem[{Baker(2022)}]{baker2022interpreting}
Baker, D.~J. (2022).
\newblock Interpreting supersymmetry.
\newblock {\em Erkenntnis\/}, {\em 87\/}(5), 2375--2396.

\bibitem[{Berghofer et~al.(2023)Berghofer, {François, J.}, {Friederich, S.},
  {Gomes, H.}, {Hetzroni, G.}, {Maas, A.}, \& {Sondenheimer,
  R.}}]{Berghofer2023}
Berghofer, P., {François, J.}, {Friederich, S.}, {Gomes, H.}, {Hetzroni, G.},
  {Maas, A.}, \& {Sondenheimer, R.} (2023).
\newblock {\em Gauge symmetries, symmetry breaking, and gauge-invariant
  approaches. Elements in the Foundations of Contemporary Physics\/}.
\newblock Cambridge University Press.

\bibitem[{Bertolini(2024)}]{bertolini2024lectures}
Bertolini, M. (2024).
\newblock Lectures on supersymmetry.
\newblock {\em Lecture notes given at SISSA\/}.

\bibitem[{Bogomolny(1976)}]{Bogomolny_eq}
Bogomolny, E.~B. (1976).
\newblock {Stability of Classical Solutions}.
\newblock {\em Sov. J. Nucl. Phys.\/}, {\em 24\/}, 449.

\bibitem[{Brading \& Castellani(2002)}]{Brading2002-BRASIP}
Brading, K., \& Castellani, E. (2002).
\newblock {\em Symmetries in Physics: Philosophical Reflections\/}.
\newblock Cambridge University Press.

\bibitem[{Butterfield(2011{\natexlab{a}})}]{butterfield2011emergence}
Butterfield, J. (2011{\natexlab{a}}).
\newblock Emergence, reduction and supervenience: A varied landscape.
\newblock {\em Foundations of physics\/}, {\em 41\/}, 920--959.

\bibitem[{Butterfield(2011{\natexlab{b}})}]{butterfield2011less}
Butterfield, J. (2011{\natexlab{b}}).
\newblock Less is different: Emergence and reduction reconciled.
\newblock {\em Foundations of physics\/}, {\em 41\/}, 1065--1135.

\bibitem[{Castellani \& Rickles(2016)}]{CastellaniRickles_Intro}
Castellani, E., \& Rickles, D. (2016).
\newblock Introduction to special issue on dualities.
\newblock {\em Studies in History and Philosophy of Science Part B: Studies in
  History and Philosophy of Modern Physics\/}, {\em 59\/}.

\bibitem[{Coleman(1985)}]{coleman1985aspects}
Coleman, S. (1985).
\newblock {\em Aspects of Symmetry\/}.
\newblock Cambridge University Press.

\bibitem[{de~Boer \& De~Haro(2004)}]{de2004off}
de~Boer, J., \& De~Haro, S. (2004).
\newblock The off-shell m5-brane and non-perturbative gauge theory.
\newblock {\em Nuclear Physics B\/}, {\em 696\/}(1-2), 174--204.

\bibitem[{De~Haro(2019)}]{deharo2019towards}
De~Haro, S. (2019).
\newblock Towards a theory of emergence for the physical sciences.
\newblock {\em European Journal for Philosophy of Science\/}, {\em 9\/}(3), 38.

\bibitem[{De~Haro \& Butterfield(2017)}]{deHaroButterfield2017}
De~Haro, S., \& Butterfield, J. (2017).
\newblock {\em A Schema for Duality, Illustrated by Bosonization\/}, (pp.
  305--376).
\newblock Springer.

\bibitem[{De~Haro \& Butterfield(2021)}]{deharobutterfield2021}
De~Haro, S., \& Butterfield, J. (2021).
\newblock On symmetry and duality.
\newblock {\em Synthese\/}, {\em 198\/}, 2973--3013.

\bibitem[{De~Haro \& Butterfield(forthcoming)}]{DeHaroButterfield2024}
De~Haro, S., \& Butterfield, J. (forthcoming).
\newblock {\em The Philosophy and Physics of Duality\/}.
\newblock Oxford University Press.

\bibitem[{Dieks et~al.(2015)Dieks, {van Dongen}, \& {De
  Haro}}]{dieksdongenharo2015}
Dieks, D., {van Dongen}, J., \& {De Haro}, S. (2015).
\newblock Emergence in holographic scenarios for gravity.
\newblock {\em Studies in History and Philosophy of Modern Physics\/}, {\em
  52\/}, 203--216.

\bibitem[{Dizadji-Bahmani et~al.(2010)Dizadji-Bahmani, Frigg, \&
  Hartmann}]{dizadji2010s}
Dizadji-Bahmani, F., Frigg, R., \& Hartmann, S. (2010).
\newblock Who’s afraid of nagelian reduction?
\newblock {\em Erkenntnis\/}, {\em 73\/}, 393--412.

\bibitem[{Franklin \& Knox(2018)}]{franklin2018emergence}
Franklin, A., \& Knox, E. (2018).
\newblock Emergence without limits: The case of phonons.
\newblock {\em Studies in History and Philosophy of Science Part B: Studies in
  History and Philosophy of Modern Physics\/}, {\em 64\/}, 68--78.

\bibitem[{Fraser(2009)}]{fraser2009quantum}
Fraser, D. (2009).
\newblock Quantum field theory: Underdetermination, inconsistency, and
  idealization.
\newblock {\em Philosophy of Science\/}, {\em 76\/}(4), 536--567.

\bibitem[{Fraser(2011)}]{fraser2011takeseriously}
Fraser, D. (2011).
\newblock How to take particle physics seriously: A further defence of
  axiomatic quantum field theory.
\newblock {\em Studies In History and Philosophy of Science Part B: Studies In
  History and Philosophy of Modern Physics\/}, {\em 42\/}(2), 126--135.

\bibitem[{Gaiotto et~al.(2013)Gaiotto, Moore, \& Neitzke}]{gaiotto2013wall}
Gaiotto, D., Moore, G.~W., \& Neitzke, A. (2013).
\newblock Wall-crossing, hitchin systems, and the wkb approximation.
\newblock {\em Advances in Mathematics\/}, {\em 234\/}, 239--403.

\bibitem[{Giveon \& Kutasov(1999)}]{giveon1999brane}
Giveon, A., \& Kutasov, D. (1999).
\newblock Brane dynamics and gauge theory.
\newblock {\em Reviews of Modern Physics\/}, {\em 71\/}(4), 983.

\bibitem[{Haag et~al.(1975)Haag, Łopuszański, \& Sohnius}]{Haag1975}
Haag, R., Łopuszański, J.~T., \& Sohnius, M. (1975).
\newblock All possible generators of supersymmetries of the s-matrix.
\newblock {\em Nuclear Physics B\/}, {\em 88\/}(2), 257--274.
\newline\urlprefix\url{https://www.sciencedirect.com/science/article/pii/0550321375902795}

\bibitem[{Hooft(1981)}]{hooft1981topology}
Hooft, G. (1981).
\newblock Topology of the gauge condition and new confinement phases in
  non-abelian gauge theories.
\newblock {\em Nuclear Physics: B\/}, {\em 190\/}(3), 455--478.

\bibitem[{Lee(2018)}]{Lee2018}
Lee, J.~M. (2018).
\newblock {\em Introduction to Riemannian Manifolds\/}.
\newblock Springer.

\bibitem[{Maas(2019)}]{Maas2019}
Maas, A. (2019).
\newblock Brout-englert-higgs physics: From foundations to phenomenology.
\newblock {\em Progress in Particle and Nuclear Physics\/}, {\em 106\/},
  132--209.

\bibitem[{Menon(2021)}]{Menon_superspace}
Menon, T. (2021).
\newblock {Taking Up Superspace: The Spacetime Setting for Supersymmetric Field
  Theory}.
\newblock In {\em {Philosophy Beyond Spacetime: Implications from Quantum
  Gravity}\/}. Oxford University Press.
\newline\urlprefix\url{https://doi.org/10.1093/oso/9780198844143.003.0005}

\bibitem[{Nagel(1961)}]{nagel1961}
Nagel, E. (1961).
\newblock {\em Te Structure of Science: Problems in the Logic of Scientific
  Explanation\/}.
\newblock Harcourt, Brace \& World, New York.

\bibitem[{Polyakov(1974)}]{Polyakov:1974}
Polyakov, A.~M. (1974).
\newblock {Particle Spectrum in Quantum Field Theory}.
\newblock {\em JETP Lett.\/}, {\em 20\/}, 194--195.

\bibitem[{Prasad \& Sommerfield(1975)}]{prasadsommerfield1975}
Prasad, M., \& Sommerfield, C.~M. (1975).
\newblock Exact classical solution for the 't hooft monopole and the julia-zee
  dyon.
\newblock {\em Physical Review Letters\/}, {\em 35\/}(12), 760.

\bibitem[{Ruetsche(2002)}]{ruetsche2002interpreting}
Ruetsche, L. (2002).
\newblock Interpreting quantum field theory.
\newblock {\em Philosophy of Science\/}, {\em 69\/}(2), 348--378.

\bibitem[{Seiberg \&
  Witten(1994{\natexlab{a}})}]{seibergWitten1994_Confinement}
Seiberg, N., \& Witten, E. (1994{\natexlab{a}}).
\newblock Electric-magnetic duality, monopole condensation, and confinement in
  $\mathcal{N}$= 2 supersymmetric yang-mills theory.
\newblock {\em Nuclear Physics B\/}, {\em 426\/}(1), 19--52.

\bibitem[{Seiberg \& Witten(1994{\natexlab{b}})}]{Seiberg_1994_symbreaking}
Seiberg, N., \& Witten, E. (1994{\natexlab{b}}).
\newblock Monopoles, duality and chiral symmetry breaking in $\mathcal{N}$ = 2
  supersymmetric {QCD}.
\newblock {\em Nuclear Physics B\/}, {\em 431\/}(3), 484--550.

\bibitem[{'t~Hooft(1974)}]{tHooft1974magneticmonopoles}
't~Hooft, G. (1974).
\newblock Magnetic monopoles in unified theories.
\newblock {\em Nucl. Phys. B\/}, {\em 79\/}(CERN-TH-1876), 276--284.

\bibitem[{Vergouwen(2022)}]{vergouwen2022emergent}
Vergouwen, S. (2022).
\newblock {\em Emergent Solitons and the Philosophy of Non-Perturbative Quantum
  Field Theory\/}.
\newblock Master's thesis, Utrecht University.
\newline\urlprefix\url{https://studenttheses.uu.nl/bitstream/handle/20.500.12932/42660/SVergouwen%20Final%20Version%20Soliton%20Thesis1.pdf?sequence=1}

\bibitem[{Wallace(2006)}]{wallace2006defence}
Wallace, D. (2006).
\newblock In defence of naivet{\'e}: The conceptual status of lagrangian
  quantum field theory.
\newblock {\em Synthese\/}, {\em 151\/}(1), 33--80.

\bibitem[{Wallace(2011)}]{wallace2011takingseriously}
Wallace, D. (2011).
\newblock Taking particle physics seriously: A critique of the algebraic
  approach to quantum field theory.
\newblock {\em Studies In History and Philosophy of Science Part B: Studies In
  History and Philosophy of Modern Physics\/}, {\em 42\/}(2), 116--125.

\bibitem[{Williams(2023)}]{Williams}
Williams, P. (2023).
\newblock {\em Philosophy of Particle Physics\/}.
\newblock Cambridge University Press.

\bibitem[{Witten(1980)}]{witten19801}
Witten, E. (1980).
\newblock The 1/n expansion in atomic and particle physics.
\newblock In {\em Recent Developments in Gauge Theories\/}, (pp. 403--419).
  Springer US.

\bibitem[{Witten(1997)}]{witten1997solutions}
Witten, E. (1997).
\newblock Solutions of four-dimensional field theories via m-theory.
\newblock {\em Nuclear Physics B\/}, {\em 500\/}(1-3), 3--42.

\bibitem[{Witten \& Olive(1978)}]{witten1978supersymmetry}
Witten, E., \& Olive, D. (1978).
\newblock Supersymmetry algebras that include topological charges.
\newblock {\em Physics Letters B\/}, {\em 78\/}(1), 97--101.

\end{thebibliography}
\end{document}